# Comment on "A First-order Phase Transition to Metallic Hydrogen"

Alexander F. Goncharov[1] and Zachary M. Geballe[1]

[1]Geophysical Laboratory, Carnegie Institution of Washington, 5251 Broad Branch Road NW, Washington D.C. 20015, USA

A recent paper by Zaghoo et al. [1] presents optical data at high-pressure and high-temperature and interprets the data as evidence for a first-order phase transition to metallic hydrogen during heating. Here we argue that the presented data are contradictory with these claims. Elucidating this issue is important for building a coherent picture that is emerging as the results of theoretical calculations of various levels [2-4] and experimental investigations employing static and dynamic compression techniques [5,6]. In this context, the use of adequate probes of the electronic and chemical state is crucial. Optical probes [1] that do not address the energy dependent conductivity while making multiple references to the Drude model are highly speculative. Indeed, the available dynamic and static compression data and theoretical modeling show that study of energy dependent conductivity is important for understanding the nature of hot dense hydrogen (see Ref. 6 and Refs. therein). Moreover, the most recent investigations [3-5], suggest that the critical point, which demarcates the regimes of crossover between insulating and plasma state at low densities to the first order liquid-liquid transition, is above 200 GPa. Below we concentrate on inconsistencies in the interpretation of data in Ref. 1, which call for careful examination of their claims and further detailed investigations. We analyze their optical data and use finite element calculations and argue that the high-temperature state studied in Zaghoo et al. is not metallic and that the data cannot discriminate between a first-order phase transition and a continuous phase transition, or even a bandgap drop within one phase.

Observations of plateaus in temperature versus heating power are the purported evidence for a first-order phase transition. In particular, Zaghoo et al. suggests that a 0.1 eV/molecule latent heat of dissociation of $H_2$ molecules is plausible and could explain the plateau. The latter argument is based on an energy balance estimate, with limited analysis of heat transport from the tungsten laser-absorber into the hydrogen medium. Our finite-element (FE) calculations of heat transport, on the other hand, show that latent heat must be orders of magnitude larger than 0.1 eV/molecule to cause plateaus similar in magnitude to those shown in the measurements of Zaghoo et al. In particular, we assume the physical properties and sample chamber geometry of Table 1, a laser pulse with a shape of Ref. 1 shown in Fig. 1, and the finite element solver of Ref. 7. We find that the latent heat of a transition at 1140 K and 170 GPa must be ~3.8 eV/molecule to reproduce the measurements of Zaghoo (Fig. 2). A latent heat of ~ 3.8 eV/molecule is implausibly high because the molecular binding energy (4.75 eV at ambient pressure), has been shown (both experimentally and theoretically) to decrease with pressure to much lower values[8,9]. We conclude that the observed plateaus are due to something besides latent heat. Fig. 2 shows that an onset of absorption at $T_c$=1140 K with a peak absorption coefficient of 0.1 µm$^{-1}$ [6] is an alternative and more plausible cause of the plateaus. The plateau emerges as a result of change in the absorption mechanism in the cavity causing a gradual time dependent growth of an absorbing hydrogen layer, reaching 270 nm in thickness in our calculations (Fig. 1). The presence of the absorbing hydrogen creates a different temperature profile in the sample cavity, with less laser power reaching the tungsten absorber and laser energy being distributed into the sample causing temperature to increase less than it would otherwise. This rearrangement results in a change in slope of measured temperature vs the laser power remarkably similar to that of Zaghoo et al. (Fig. 2). This is the first time these measurements have been quantitatively modelled based on physical principles, and rule out previous qualitative interpretations.

Decreases in optical transmission by 3% to 20% and increases in reflection of up to 13% in the plateau region are interpreted as being due to transformation of thin layers of hydrogen to a metallic state. In particular, Zaghoo et al. estimates electrical conductivity of 2100 S/cm at 1250 K and 170 GPa. A 10 nm thickness of the transformed hydrogen is assumed. However, our heat-flow models show that the thickness of transformed hydrogen reaches ~ 200 nm when the peak temperature exceeds the reported transition temperature by ~ 100 K (as in the data used to estimate electrical conductivity in Ref. 1). A 10 nm thick sample can be only obtained in the regime of very small overheating of approximately 2 K above the $T_c$, which is experimentally not achievable. We present in Fig. 3 the Drude model that matches the observed reflectivity (13%) and assumes a "full dissociation" of hydrogen molecules but strongly disagrees in transmission value (5% vs 93% measured) requiring the sample to be only 4.9 nm thick to match the reported transmission. We find that the Drude model parameters of Fig. 3 yield the DC conductivity of 590 S/cm suggesting semiconductor or bad metal behavior. The electrical conductivity is even lower if non-Drude, semi-metal optical properties are accounted for[6].

In summary, we refute the claim of Zaghoo et al. that "Our pulses are carefully tailored to have just sufficient energy to metallize a thin film of hydrogen". Rather, relatively large increases in heating power and in peak temperature are needed to build up a layer of absorbing hydrogen that is measurable, meaning that a relatively thick layer (~200 nm) is created. The reported plateau is unlikely due to latent heat, as even such thick layer does not produce an anomaly in temperature versus the laser power that would be consistent with plausible latent heat values. The optical properties of the layer of transformed hydrogen are inconsistent with those of metal as reflectivity is too small and transmission is too large, making the claims of metallization premature. Hence, we find no reliable evidence for the adjectives in the paper's title; the high-temperature hydrogen is not metallic and the detected transformation need not be a phase transition, much less a first-order phase transition. On the other hand, Zaghoo et al. does show interesting changes in the optical properties of hydrogen that warrant further study.

We thank Stewart McWilliams for valuable comments and discussions.

Table 1. Material properties and thickness of layers assumed in our finite element calculations at 170 GPa.

|  | Hydrogen | Alumina | Diamond | Tungsten |
|---|---|---|---|---|
| Thermal conductivity (W/m/K) | 100 | 100 | 2000 | 226 |
| Specific heat capacity (J/kg/K) | 15000 | 880 | 509 | 134 |
| Density (kg/m$^3$) | 772 | 5500 | 3500 | 30000 |
| Layer thickness (μm) | 0.270 | 0.05 | 30 | 0.01 |

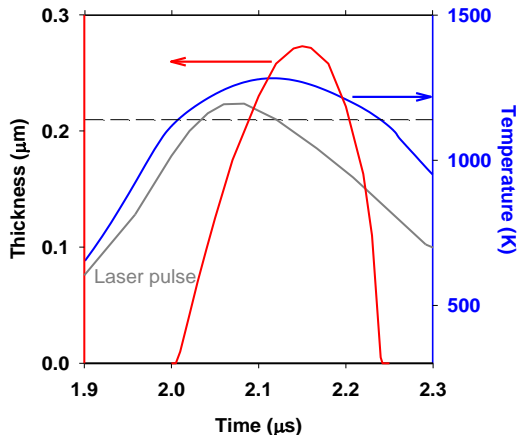

**Fig. 1.** The calculated transformed sample thickness heated to the maximum temperature of 1280 K as in experiments of Ref. 1. The transformation temperature, $T_c$=1140 K, above which hydrogen absorbs is shown by a horizonatal dashed line.

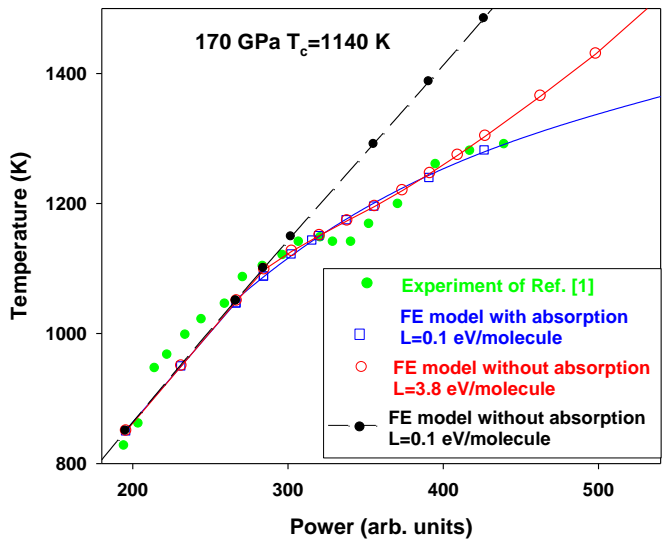

**Fig. 2.** The sample temperature vs the laser power measured radiometrically in Ref. 1 compared to FE calculations that assume that hydrogen absorbes above $T_c$=1140 K and with no absorption. The calculations take into account the variable latent heat (L) associated with the transition at $T_c$ (broadened by a 100 K using the Gauss error function).

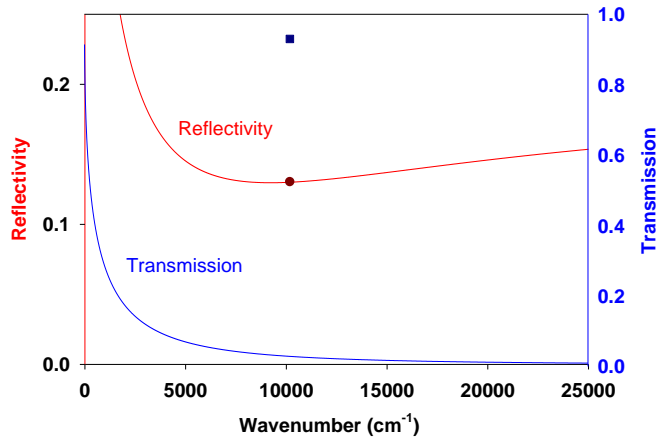

Fig. 3. The optical properties of of hot hydrogen calculated using the Drude model with the following parameters: $W_p$=14.4 eV, scatterung time $\tau=1.4*10^{-17}$ s. The sample thickness is 200 nm. The solid points (the circle- reflectivity and the square- transmission) are from Ref. 1 for 980 nm, 170 GPa, T=1280 K.